\documentclass[a4paper]{spie}  
\usepackage{graphicx}
\usepackage{amssymb, amsmath, subfigure}

\title{Fluctuations and noise in cancer development.} 

\author{Matthew J. Berryman\supit{a}, Sabrina L. Spencer\supit{a,b}, Andrew Allison\supit{a}, and Derek Abbott\supit{a}
\skiplinehalf
\supit{a}Centre for Biomedical Engineering and\\
School of Electrical and Electronic Engineering,\\
The University of Adelaide, SA  5005, Australia.\\
\supit{b}Department of Human Genetics,\\
University of Michigan,\\
Ann Arbor, MI  48109, USA.}
\authorinfo{Send correspondence to Derek Abbott\\
E-mail: dabbott@eleceng.adelaide.edu.au, Telephone: +61 8 8303 5748}

\begin{document} 
\maketitle
\begin{abstract}
This paper explores fluctuations and noise in various facets of cancer development. The three areas of particular focus are the stochastic progression of cells to cancer, fluctuations of the tumor size during treatment, and noise in cancer cell signalling. We explore the stochastic dynamics of tumor growth and response to treatment using a Markov model, and fluctutions in tumor size in response to treatment using partial differential equations. We also explore noise within gene networks in cancer cells, and noise in inter-cell signalling.
\end{abstract}
\keywords{Inter-cellular signalling, cancer treatment, tumor growth, gene pathways, gene networks.}
\section{INTRODUCTION}
Fluctuations and noise are present in a wide variety of physical systems~\cite{Ginzburg03,Belyakov03,Telesca02}. This paper explores the effects of noise in tumor growth and response to treatment, and in general ``noise'' in gene networks and in inter-cell signaling. This improves our understanding of the nature of cancer development and growth, through its examination of the robustness and sensitivity of the gene pathways that are important in the progression of normal cells to cancerous cells. 
\section{METHODS}
For both the tumor response and Wnt signaling analysis, the ODEs are solved using the Runge-Kutta method of order 5 with variable step size. We note that although an Euler method has been shown to work better than the Runge-Kutta method, it has been shown that using the Runge-Kutta method of order 4 with step sizes much greater than those used here (of a maximum $h=0.1$) works well for both the linear additive noise used in the Wnt signaling modeling and in particular for the multiplicative noise used in the tumor response~\cite{Davis}.

We explore three models useful for studying cancer. Two models consider the effect of tumor treatment, one a Markov model and the other an ODE model. In the Markov, population-based model, we consider stochastic effects on tumor growth when treatment is applied. Observing the fluctuations in the response leads us to consider adding such fluctuations to an existing ODE model of tumor treatment~\cite{tjacks}, which considers the cells in regions of 3D space. We also explore a model of inter-cell signalling and the effect of adding noise to it, as inter-cell signaling plays an important role in the development of cancer~\cite{betacatenin1}.
\subsection{Tumor treatment using Markov model}
We consider in our models the following cell populations: normal epithelial cells denoted by $N$ cells with mutations that lead to genetic instability ($G$), cells with mutations which increase their replication rate ($R$), cells with mutations which allow them to avoid death ($D$), cells which have acquired the ability to induce angiogenesis ($A$), and cells with two or more mutations. Cell populations that have acquired two or three mutations are denoted by listing the mutations together. We label a cell which has acquired all four mutations a tumor cell ($T$). Finally, a tumor cell which has acquired the capability to invade and metastasize is a cancer cell ($M$).

The spontaneous mutation rate in human cells has been estimated to be in the range of $10^{-7}$ to $10^{-6}$ mutations/gene/cell division~\cite{jackson98}. We assume a mutation rate of $k_{1}=10^{-6}$/gene/cell division. As genetic instability favors the accumulation of mutations~\cite{cahill99}, we assume that the mutation rate after a genetic instability mutation increases to $k_{2}=10^{-4}$ mutations/gene/cell division~\cite{Nowak5}. Successful invasion and metastasis depend upon acquisition of the other capabilities, as well as several new capabilities~\cite{hanahan00}.  To simplify the model, we do not address the multistep progression of a tumor cell to a metastatic cell, and instead consider this complex process as one step. It is estimated that only $1$ in $10^{9}$ cancer cells will invade and successfully metastasize to a new site~\cite{mool}, and so we use a rate of $k_{3}=10^{-9}$ for the transition from a tumor cell to a metastatic cell.
In addition, our model assumes that a tumor cannot grow past $\approx10^{6}$ cells without angiogenesis supplying blood to the tumor~\cite{Folkman}, and thus we cap the size of the tumor at $C_{1} = 10^{6}$ cells until the tumor acquires a mutation in an $A$ gene. Death of the patient occurs when the tumor reaches $C_{2} = 10^{12}$ cells~\cite{surgonc}.  The Markov model also takes into account the fact that there are approximately $100$ genes involved in each category of mutations~\cite{Nowak5}. 

The observed volume doubling time of tumors in clinical practice is on the order of $100$ days~\cite{divisions}.
We assume that the relative contribution to increased net proliferation for mutations in the $D$ and $R$ categories is $0.8$ and $0.2$,
 respectively.  This ratio places more importance on apoptosis mutations than the $0.7:0.3$ inferred from a paper by Tomlinson and Bodmer~\cite{tomlinson2} due to the crucial role of apoptosis in net cell proliferation~\cite{igney}. There is no net proliferation associated with cells which do not have mutations in $D$ or $R$. The net doubling times associated with $D$ and $R$ in our models are $500$ and $200$ days, respectively. Cells with a mutation in $R$ and $D$ have a net doubling time of $100$ 
days. Cells without a mutation in an $R$ gene divide every $b=5$ days; cells with a mutation in an $R$ gene divide every $b_{R}= 4.96$ days. The cell lifetime used for cells without a mutation in a $D$ gene is $d=5$ days; the lifetime used for cells with a mutation in a $D$ gene is $d_{D} = 5.08$ days. 
The birth and death rates are equal for normal cells, thus there is no net proliferation unless a mutation in $D$ or $R$ is acquired.  The cell division rates are conservative estimates based on Rew and Wilson~\cite{divisions}.

Our model makes several simplifications. In the case where some cells might not have a mutation in $A$ but neighboring cells do, we still limit the population of cells without mutations in $A$ even though they may have an adequate blood supply due to signals from neighboring cells. In addition, we do not take into account the fact that mutations in some genes may fall in two or more categories.  For example, a mutation in p53 may fall into both the $D$ and $G$ categories, as p53 is involved in both apoptotic and DNA repair pathways~\cite{surfing}.

\begin{figure}[htb]
\centering
\includegraphics[scale=0.35]{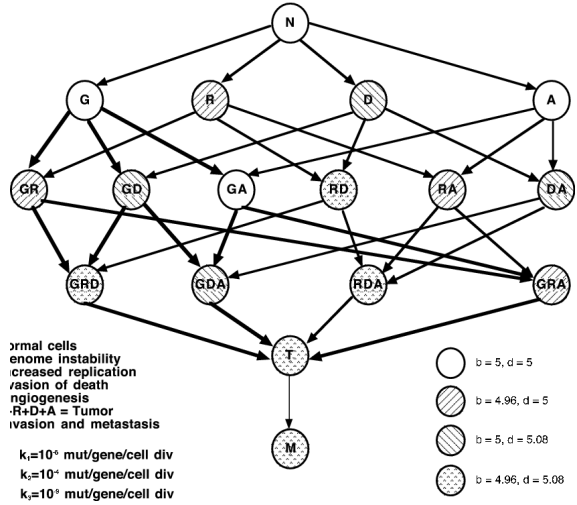}
\caption{State diagram of the model. Normal cells $(N)$ can acquire mutations which lead to genetic instability, $(G)$; mutations which increase the proliferation rate, $(R)$; mutations which give the cell the capability to avoid death, $(D)$; or mutations which give the cell the capability to induce angiogenesis, $(A)$. These mutations are acquired at rate $k_{1}$. After a mutation in $G$, the mutation rate increases to $k_{2}$. When a cell has acquired a mutation in each category, it becomes a tumor cell $(T)$. Finally, tumor cells become metastatic cells $(M)$ at rate $k_{3}$.}
\label{model}
\end{figure}

We consider a discrete time Markov model, using the states and transition rates as shown in Figure~\ref{model}. We use state-dependent time steps of $\Delta t = 1/b$, where $b$ is the cell division time, as we consider the mutations to become fixed at each cell division. For example, if a cell is in the $R$ state, each $b_{R}$ days, the cells gain mutations with probability
\begin{equation}
P=p_{s}(t)k_{s}g,
\label{transprob}
\end{equation}
where $p_{s}(t)$ is the population of cells in state $s$ at time $t$, $k_{s}$ is the rate of transition out of state $s$, and $g$ is the number of genes involved in each state, which we keep constant at $100$.
We begin with a population of normal cells (except when we are modeling inherited mutations). We then allow the cells to mutate through a particular path of states, dividing at a rate which depends on the current state. The net proliferation depends on the current state as well. We treat the total population of cells as a homogeneous population, and run the simulation for a large number of trials ($10,000$). We then average in order to simulate a heterogeneous population.
\subsection{Tumor treatment using an ODE model}
We use a partial differential equation based model for  tumor size based on work by Jackson and Byrne~\cite{tjacks}. The model treats the tumor as a spherical mass with rapidly dividing, highly drug-susceptible cells, surrounded by normal cells that have a lower response to the drug treatment.  The parameters used are defined in Table~\ref{paramtable}, and the symbols in Table~\ref{symtable}, as used in Equations~\ref{db}-\ref{R}.
\begin{table}[htbp]
\centering
\caption{List of the parameters used in the model and the values used from the literature~\cite{tjacks,tjacks2}. The parameter $\Gamma$ is the rate coefficient of blood-tissue transfer, or the rate at which the drug can pass from the blood into cells. The parameter $\beta$ is the association rate, or binding of drug to target receptors in the tumor. The rates of drug loss due to decay, molecular instability, or cellular uptake and metabolism are combined in the parameter $\lambda$. Transfer coefficients $k_{12}$ and $k_{21}$ are used to describe transfer of the drug out of and into the group of rapidly dividing, highly responsive cells, respectively. Parameter $k_{e}$ represents elimination of the drug from blood plasma.}
\begin{tabular}{|c|c|c|}\hline
{\bf parameter} & {\bf value}\\\hline
$\Gamma$ & 16 day$^{-1}$\\\hline
$\beta$ & 560 M$^{-1}$day$^{-1}$\\\hline
$\lambda$ & 1.9 day$^{-1}$\\\hline
$k_{1}$ & $2.4\times10^{-5}$ s$^{-1}$\\\hline
$k_{2}$ &  $5.8\times10^{-4}$ s$^{-1}$\\\hline
$k_{e}$ & $7.1\times 10^{-5}$ s$^{-1}$\\\hline
\end{tabular}
\label{paramtable}
\end{table}
\begin{table}[htbp]
\centering
\caption{Variables used in the model.}
\begin{tabular}{|c|c|}\hline
{\bf symbol} & {\bf description}\\\hline
$d_{B}$ & prescribed drug concentration in the tumor vasculature (M)\\\hline
$d_{N}$ & the drug concentration in the surrounding tissue (M)\\\hline
$R$ & the radius of the tumor (mm)\\\hline
\end{tabular}
\label{symtable}
\end{table}
\begin{equation}
\frac{d_{B}}{dt}=-k_{12}d_{B}+k_{21}d_{N}-k_{e}d_{B},
\label{db}
\end{equation}
\begin{equation}
\frac{d_{N}}{dt}=k_{12}d_{B}-k_{21}d_{N},
\label{dn}
\end{equation}
\begin{equation}
\frac{dR}{dt}=\left(1-\frac{\beta\Gamma}{\xi^{2}}d_{B}(t)\right)\frac{R}{3}
-\frac{\beta}{\xi}\left(d_{N}(t)-\frac{\Gamma}{\xi}d_{B}(t)\right)
\frac{\xi R \mathrm{cosh} \xi R - \mathrm{sinh} \xi R}{\xi R \mathrm{sinh} \xi R},
\label{R}
\end{equation}
where $k_{e}$, $k_{12}$, and $k_{21}$ are given in Table~\ref{paramtable}, and $\xi=\sqrt{\Gamma+\lambda}$.
The initial condition for $d_{N}$, $C$, is given by
\begin{equation}
C=\sqrt{k_{12}^{2}+2k_{12}k_{e}+2k_{12}k_{21}+k_{e}^{2}+k_{21}^{2}/k_{12}},
\label{C}
\end{equation}
where $k_{12}$, $k_{21}$, and $k_{e}$ are as defined in Table~\ref{paramtable}.
The initial conditions used are $R(0)=0.4$ cm$^{-1}$ (in line with Kerr {\it et al.}~\cite{Kerr}), $d_{B}(0)=0.1$, and $d_{N}=C$.
We consider fluctuations of the tumor size, $R$, in the response to the drug concentration in the tumor, $d_{B}$. We multiply an exponential noise term $\alpha$, from an exponential distribution $P(x)=\lambda e^{-\lambda x}$, by $d_{B}$ to show fluctuations in tumor response in short intervals, and decreased tumor responsiveness to treatment over time. Tumor responsiveness may vary with fluctuations in cell signaling, which we explore in the following subsection.
\subsection{Inter-cell signalling}
The Wnt signaling pathway contains many oncogenes~\cite{Polakis00}. Although the evidence for mutations in and overexpression or underexpression of Wnt in development of cancer is not entirely conclusive~\cite{Polakis00,WntBreastCancer}, there is growing evidence that the downstream targets such as APC and $\beta$-catenin play an important role in the development of many cancers~\cite{WntBreastCancer,Polakis00,betacatenin1}.

\begin{figure}[hbtp]
\centering{\resizebox{10cm}{!}{\includegraphics{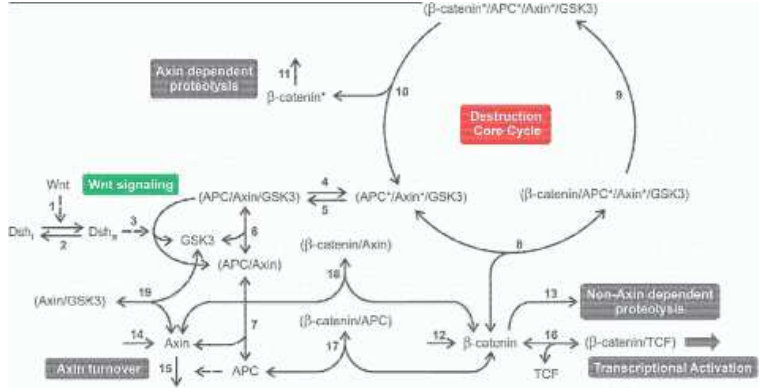}}}
  \caption{This figure (from Lee {\it et al.}~\cite{WntMath}) shows the $\beta$-degradation cycle, and the protein interactions are shown with arrows.}
\label{ReactionScheme}
\end{figure}

Here we consider the reaction scheme for Wnt signaling shown in Figure~\ref{ReactionScheme}. 
In this reaction scheme, Wnt, through binding to its receptor Frizzled, activates Dsh. This activated Dsh protein then binds to axin, and through bound GBP, inhibits the phosphorylation of $\beta$-catenin. 
Axin plays a key role in coordinating the assembly of large protein complexes including one that forms from glycogen synthase kinase 3$\beta$ (GSK3$\beta$), adenomatous polypopsis coli (APC) and the the negateive regulators Dsh and GSK3$\beta$-binding protein (GBP). The absence of Wnt signals for the phosphorylation of $\beta$-catenin, marking it for subsequent degradation in the degradation cycle shown. Since $\beta$-catenin plays a role in signaling events in the cell-cycle, the Wnt pathway plays a role in tumor formation, where the cell-cycle becomes disrupted~\cite{hanahan00}.

The Wnt signaling pathway can be modeled using a set of ordinary differential equations~\cite{WntMath}. We use the same set of equations in modeling $\beta$-catenin degredation, however we expand the range of functions used to model the transient stimulation of Wnt {\it in vitro}, and model various types of noise on these inputs.

Although the degradation of Wnt is assumed to be an exponentially decaying function in Lee {\it et al.}~\cite{WntMath}, this assumes that:
\begin{itemize}
\item the degradation process is random, 
\item independent of time, 
\item has a constant probability per unit time for each Wnt molecule.
\end{itemize}
It is not obvious biologically that the process is entirely random, nor that the probability per unit time is entirely constant. While clearly the end result of a decrease in Wnt signalling is an exponential-like decay in $\beta$-catenin~\cite{WntMath,Aberle97}, there are many simpler expressions which also give such a decay in $\beta$-catenin. 

The various functions for the Wnt degradation, $W\left(t\right)$, we use are given in Equations~\ref{wnt:one} through~\ref{wnt:three}. First, we consider the simplest function, that the Wnt is degraded linearly:
\begin{equation}
W\left(t\right)=
\begin{cases}
n\left(t\right), & t<t_{0},\\
\mathrm{max}\left\{1-\lambda\left(t-t_{0}\right),0\right\}+n(t), & t > t_{0},
\end{cases}
\label{wnt:one}
\end{equation}
where $\lambda=1/\tau_{\scriptstyle W}$ for $\tau_{\scriptstyle W}$ the characteristic lifetime of 20 minutes for Wnt, $t_{0}$ is the time at which the signaling occurs, and $n(t)$ is the noise term described later. This linear model changes the assumption of constant probability per unit time for each molecule; instead it describes a constant rate of breakdown independent of the amount of Wnt. In other words, this represents saturation of Wnt receptors.
We also tried the same exponential decay equation as used by Lee {\it et al.}, however with a noise term, $n(t)$, added:
\begin{equation}
W\left(t\right)=
\begin{cases}
0, & t<t_{0},\\
e^{-\lambda \left(t-t_{0}\right)}+n\left(t\right) & t>t_{0},
\end{cases}
\label{wnt:two}
\end{equation}
We used a combination of Equations~\ref{wnt:one} and~\ref{wnt:two}, to reflect saturation of Wnt receptors, as in Equation~\ref{wnt:one}, to the point where it can be degraded in an independent fashion as in Equation~\ref{wnt:two}:
\begin{equation}
W\left(t\right)=
\begin{cases}
n\left(t\right), &t<t_{0},\\
1-\lambda\left(t-t_{0}\right)+n\left(t\right), & t_{0} < t < 1/\lambda+W_{c}/\lambda+t_{0},\\
W_{c}e^{-\lambda \left(t-1/\lambda+W_{c}/\lambda\right)}+n\left(t\right), & t>1/\lambda+W_{c}/\lambda+t_{0}/\lambda,
\end{cases}
\label{wnt:three}
\end{equation}
Where $0<W_{c}<1$ is the level at which Wnt can be degraded in an independent fashion.
The types of noise, $n(t)$ we consider are:
\begin{itemize}
\item uniform noise in the range $(0,0.1)$, which represents a small, approximately constant background level of Wnt over time,
\item Poisson noise, representing a a level of Wnt which occasionally fluctuates to high levels through time, and
\item exponential noise, similar to the Poisson noise but with no structure to the time between fluctuations.
\end{itemize}
\section{RESULTS}
\subsection{Tumor treatment using a Markov model}
\begin{figure}[htbp]
\centering
  \mbox{
  \subfigure[]{
  \includegraphics[height=6cm]{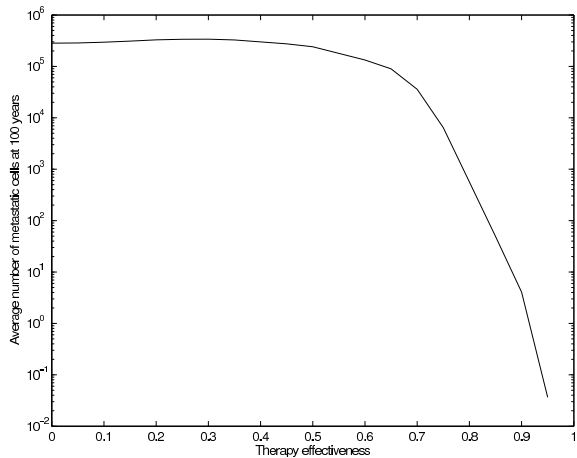}
  \label{therapy2:b}}}
  \\\mbox{
  \subfigure[]{
  \includegraphics[height=6cm]{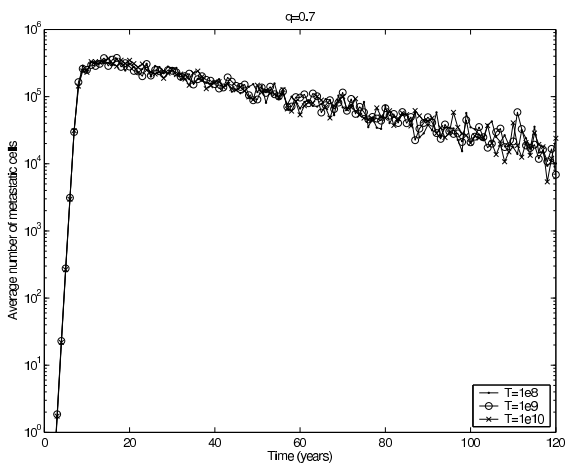}
  \label{therapy2:c}}}
  \mbox{
  \subfigure[]{
  \includegraphics[height=6cm]{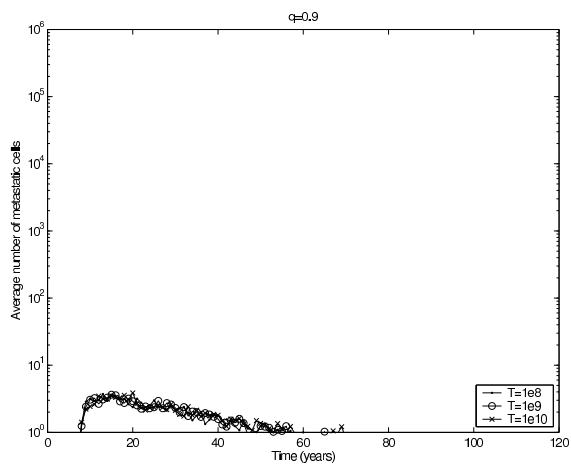}
  \label{therapy2:d}}}
\caption{Simulations of therapy using the Markov model with varying parameters of effectiveness and cell population.
 (a) Number of metastatic cells at 100 years after applying surgery at $10^{9}$ tumor cells followed by chemotherapy with varying degrees of effectiveness.
 (b) Effect of therapy that is $70\%$ effective, applied at different tumor sizes: $10^{8}$, $10^{9}$, and $10^{10}$.
 (c) Effect of therapy that is $90\%$ effective, applied at different tumor sizes.}
\label{therapy2}
\end{figure}

Figure~\ref{therapy2} shows the effect of applying surgery followed by 
intermittent therapy to tumor masses of various sizes. The 
stochastic nature of the Markov model is visible, although use of a 
large number of simulations ($10,000$) reduces this somewhat. 

We show a noticeable drop in cell population in Figure~\ref{therapy2:b} as therapy effectiveness, the percent of cells being killed on replication, increases from  70\% to 90\%. The minimum effectiveness to significantly reduce the population of metastatic cells is reasonably realistic here as the Markov model captures the discrete time nature of the intermittent treatment. The transition from very little effect to a significant effect on the population of metastatic cells is fairly gradual between 70\% and 90\%.
The effect of this increase in treatment effectiveness on cell population is shown in Figures~\ref{therapy2:c} and~\ref{therapy2:d}, where we plot the number of $M$ cells over time with varying tumor sizes at the start of treatment.

From Figures~\ref{therapy2:c} and~\ref{therapy2:d}, it is clear that the treatment effectiveness is more important than the tumor size at which therapy is applied. With a therapy effectiveness of 70\%, there are $\approx 10^{4}$ $M$ cells at 100 years on average, and with a therapy effectiveness of 90\% there are no $M$ cells at 100 years on average. 

The level of fluctuations in tumor size increase over time, highlighting the variation in tumor response to treatment as some cell populations within the heterogeneous tumor mass gain resistance to the treatment drugs used. This is one reason why we consider such variations in the tumor response ODE  model.
 \subsection{Tumor treatment using a DE model}
\begin{figure}[htbp]
\centering\mbox{
		\subfigure[Growth of tumor with treatment, no fluctuations in response.]{
	\label{tumorkill:check}
	\includegraphics[width=7cm]{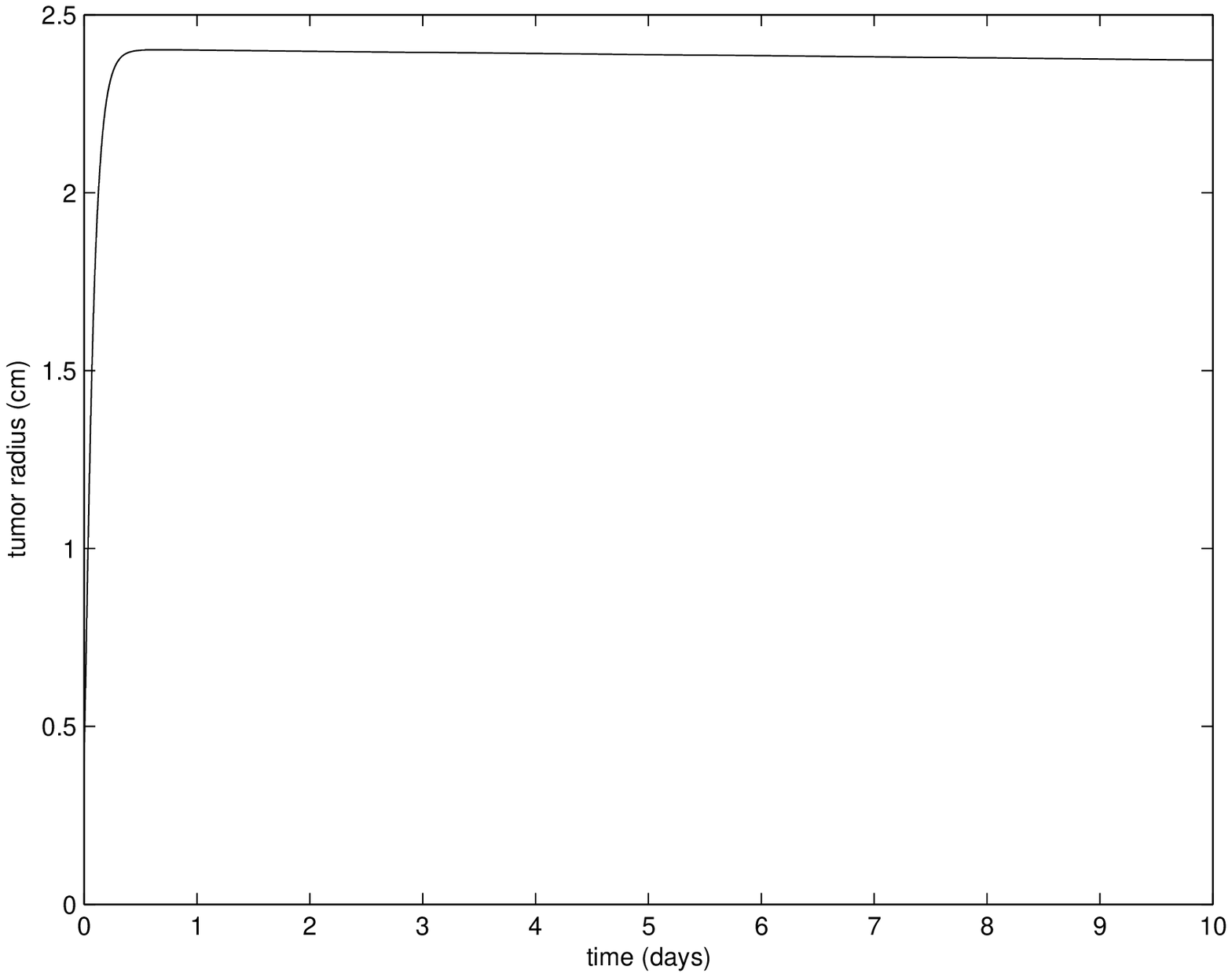}
	}}\mbox{
	\subfigure[Growth of tumor with treatment, fluctuations in response.]{
	\label{tumorkill:uncheck}
	\includegraphics[width=7cm]{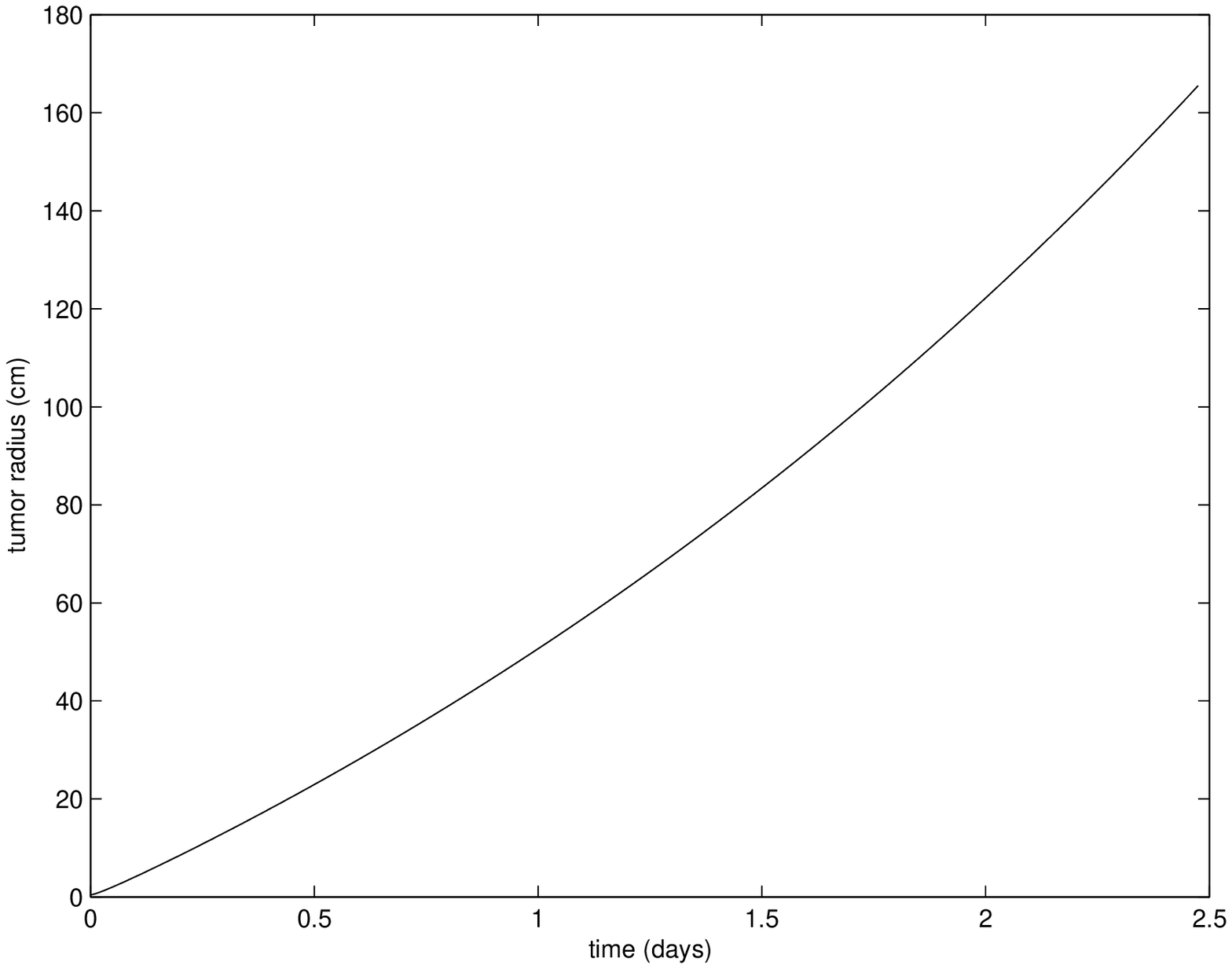}
	}}
  \caption{ODE model of tumor growth and response. This figure demonstrates the controlled tumor growth for the case where treatment is applied with no fluctuations in tumor response (a) and uncontrolled tumor growth for treatment with minor fluctuations (b). In (a), the tumor initially grows in size up to the point where the surface area is large enough that the drug can enter the tumor in high enough concentrations to start having an effect. Note the difference in scale of the $y$-axis between the two graphs.}
\label{tumorkill}
\end{figure}

Normally, application of the treatment holds the growth of the tumor in check, as shown in Figure~\ref{tumorkill:check}. The tumor expands to a size at which a large amount of the drug can enter, then starts to shrink reasonably slowly. Multiplying the tumor response to treatment term, $d_{B}(t)$ by exponential noise, with probability distribution $P(x)=\lambda e^{- \lambda x}$,
 for a wide variation in $\lambda$ (0.5 through 1.5), produces uncontrolled tumor growth within a very short period of time, as shown in Figure~\ref{tumorkill:uncheck}. Our results highlight the importance of considering fluctuations in the tumor response. The difference they make to the model is stark: we have gone from a very controlled, good response to treatment to a situation where the tumor grows uncontrollably. Such fluctuations may be caused by variations in cell division rate (as shown by the Markov model) which affects the response to the drug, or other cellular responses such as inter- and intra-cell signaling.
\subsection{Inter-cell signalling}
\begin{figure}[htbp]
\centering\mbox{
		\subfigure[This shows the linear, mixed, and exponential degradation of the Wnt input signal over time, with no noise added.]{
	\label{nonoise:input}
	\includegraphics[width=7cm]{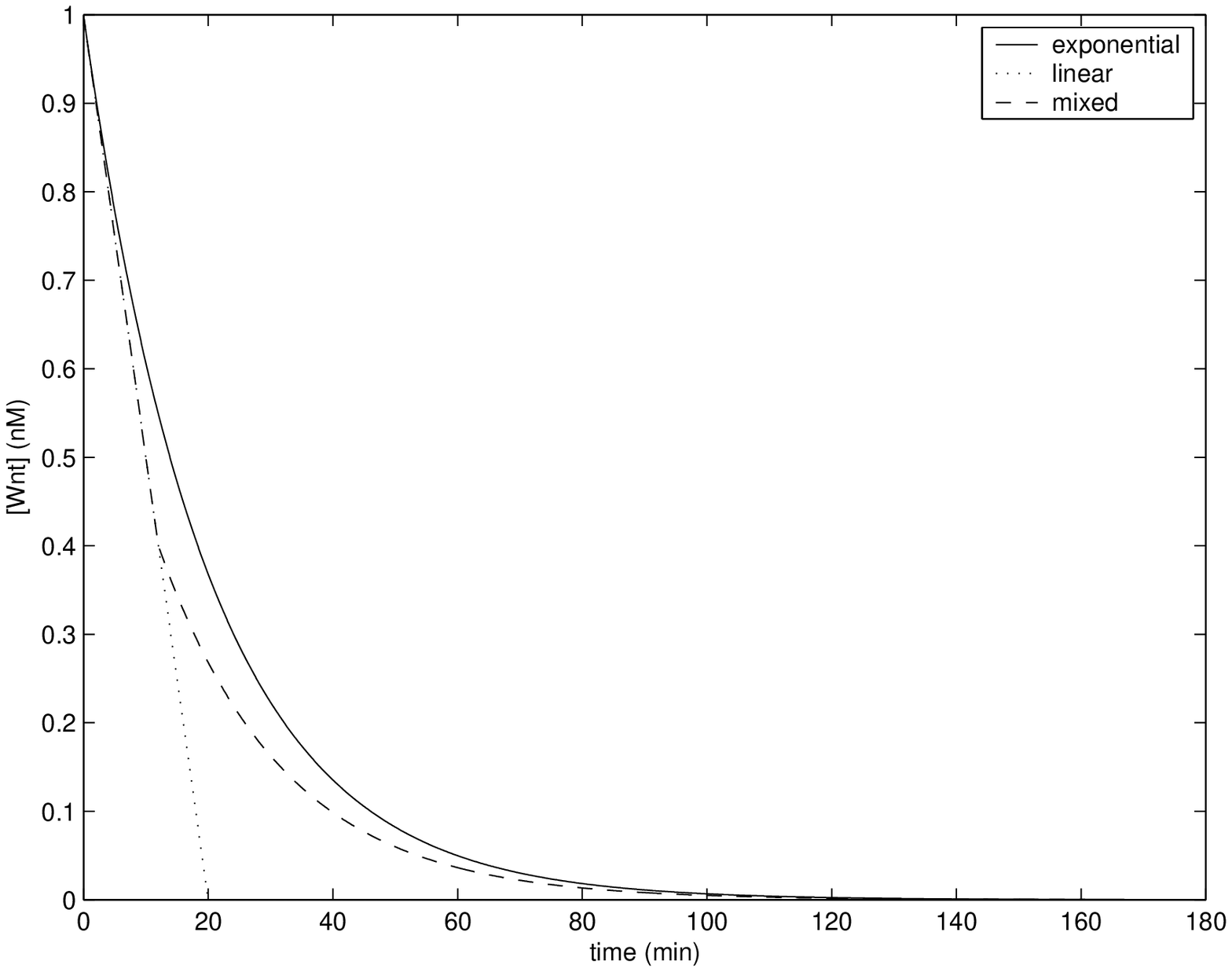}
	}}\mbox{
	\subfigure[Here we show the response of the system for varying inputs. We plot the concentration of $\beta$-catenin over time for the exponential, linear, and mixed linear and exponential decrease in the level of Wnt. ]{
	\label{nonoise:output}
	\includegraphics[width=7cm]{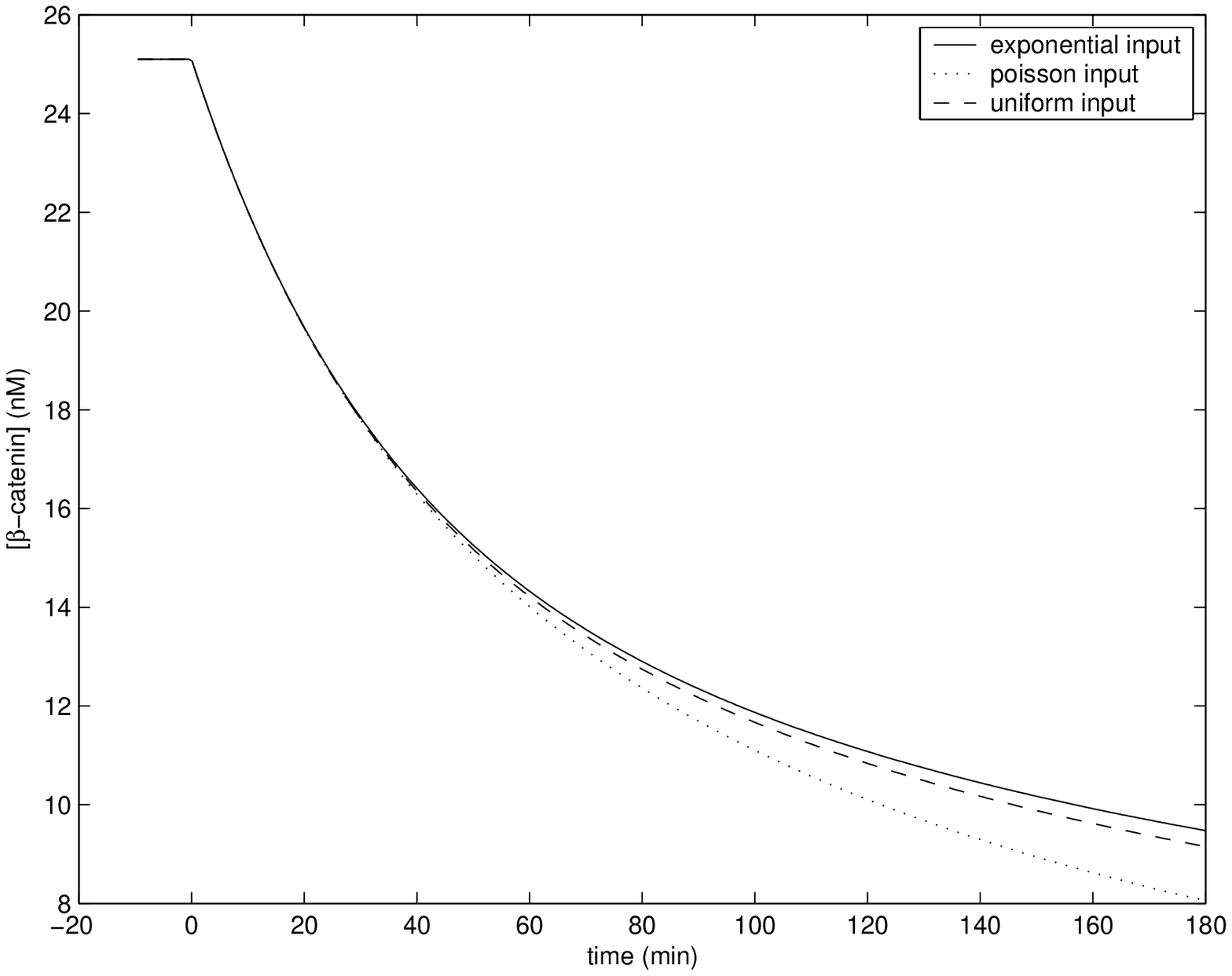}
	}}
  \caption{Input and output of the Wnt signaling system in the absence of noise. Note that all inputs give exponential decays of $\beta$-catenin, however a linear decrease in Wnt causes a larger decrease in $\beta$-catenin. This is due to the fact that for the linear decrease, the Wnt receptors are being saturated and thus signal for degradation of $\beta$-catenin at a higher rate. The mixed model appears somewhere between the two, as the Wnt receptors are only initially saturated.}
\label{nonoise}
\end{figure}

\begin{figure}[htbp]
 \centering\mbox{
        	\subfigure[]{
            \label{noise:exp}
            \includegraphics[width=6.5cm]{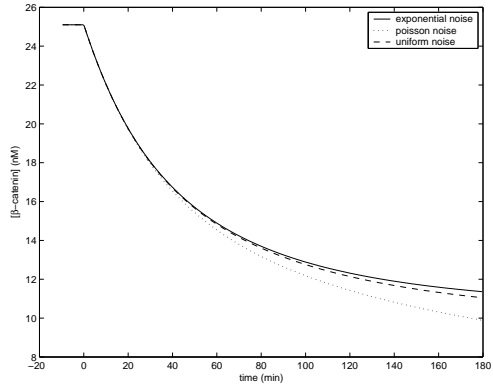}
        }}\\\mbox{
        \subfigure[]{
            \label{noise:lin}
            \includegraphics[width=6.5cm]{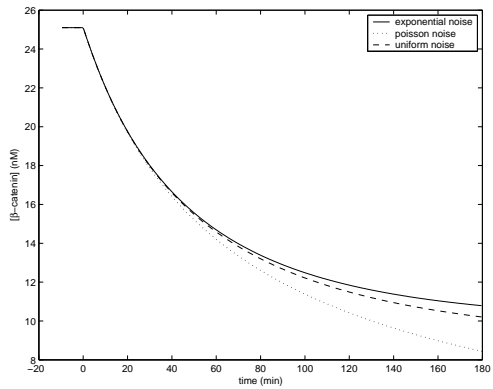}
       }}\mbox{
        \subfigure[]{
            \label{noise:mixed}
            \includegraphics[width=6.5cm]{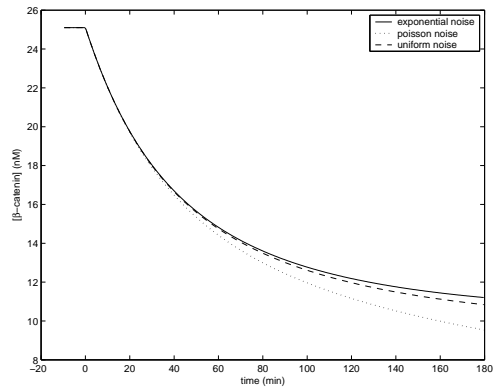}
        }}
        \caption{These figures show the result of adding uniform, Poisson, and exponential noise to (a) exponentially decaying Wnt input, (b) linearly decaying input, and (c) exponential followed by linear decaying input. We add noise that is uniform on $(0,0.1)$, Poisson with mean $0.1$, or exponential with a mean of $0.1$.}
\label{noise}
\end{figure}

The results of using the linear, exponential, and mixed rates of Wnt decay as per Equations~\ref{wnt:one}-\ref{wnt:three}, and with no noise added, are shown in Figure~\ref{nonoise}. All the various forms of noise added to the system increased the level of $\beta$-catenin at 180 minutes, with exponential noise increasing it the most. The values are shown in Table~\ref{endvalues}. For uniform and exponential (correlated) noise there is a higher overall background level of Wnt, which translates into a higher  level of $\beta$-catenin in the system, when compared with Poisson (uncorrelated) noise.
\begin{table}
\centering
\caption{Here we show the level of $\beta$-catenin (in nM) remaining 180 minutes after removal of steady-state Wnt signal, for all three types of decaying input signals and with no noise and all three types of noise added.}
\begin{tabular}{|c||c|c|c|}\hline
 & Exponential input & Linear input & Mixed input\\\hline\hline
No noise & 9.48 & 8.07 & 9.15 \\\hline
Exponential noise & 11.36 & 10.79 & 11.21\\\hline
Uniform noise & 11.05 & 10.21 & 10.84\\\hline 
Poisson noise & 9.89 & 8.43 & 9.53\\\hline
\end{tabular}
\label{endvalues}
\end{table}
We then added in the set of noise terms, $n(t)$ into Equations~\ref{wnt:one}-\ref{wnt:three}. The result of adding in noise is shown in Figure~\ref{noise}.
\clearpage
\section{CONCLUSIONS}
It is difficult to draw conclusions from model predictions of response in the absence of noise (here used to model fluctuations in tumor response). While it is difficult to extrapolate models of tumor growth to {\it in vivo} outcomes due to inability to detect small tumors~\cite{surgonc}, the dramatic response of the ODE model to the addition of noise highlights the need to build better, stochastic models of tumor growth. The Markov model also shows how the response of the tumor fluctuates with time, which impacts on the predicted level of growth while undergoing treatment.
Both the ODE and Markov models suggest a large tumor response to treatment is needed to dramatically improve the effectiveness of treatment.

In modeling inter-cell signaling, omitting the noise terms results in lower than predicted $\beta$-catenin levels. Given this, it is important to consider including noise when modeling {\it in vivo} $\beta$-catenin; even small changes in Wnt-mediated $\beta$-catenin expression result in clinically observable changes in tumor response~\cite{betacatenin1,betacatenin2}.

The models used are quite limited in terms of accurate modeling of the tumor growth and response, and the Wnt signaling. They make several broad assumptions, in particular the ODE model of tumor response assumes a homogeneous spherical mass and neglects the inner mass of necrotic cells surrounded by an irregular-shaped mass of rapidly dividing cells as dealt with in other models~\cite{Mansury}. Nonetheless they make useful predictions about cell-cell interactions and cell interactions with chemotherapy drugs. Quantitatively, adding noise to the tumor response in both the Markov and ODE models reduces the effectiveness in treatment; this is markedly shown in the ODE model. Adding noise to cell-cell interactions changes the levels of protein expression in the cell, which can then lead to an effect on tumor response. Building more accurate models would allow us to make more useful predictions of the effects of chemotherapy drugs. We propose further tests of these models in {\it in vitro} and {\it in vivo} settings, allowing further refinement of the models.
\acknowledgments
We are very appreciative of the support and advice provided by Dr.~Gilbert Omenn and Dr.~Trachette Jackson of the University of Michigan. We acknowledge the funding provided by The University of Adelaide.
\bibliography{phd}   
\bibliographystyle{spiebib}   
\end{document}